\documentclass[aps,prl,reprint,groupedaddress]{revtex4-1}

\usepackage{graphicx}
\usepackage{dcolumn}
\usepackage{bm}
\usepackage{amssymb}
\usepackage{color}
\begin{document}


\title{Intermediate Scaling and Logarithmic Invariance in Turbulent Pipe Flow}

\author{Sourabh S. Diwan}
\email[]{sdiwan@iisc.ac.in}
\thanks{Present address: Department of Aerospace Engineering, Indian Institute of Science, Bangalore 560012.}
\affiliation{Department of Aeronautics, Imperial College London, South Kensington, London SW7 2AZ.}

\author{Jonathan F. Morrison}
\email[]{j.morrison@imperial.ac.uk}
\affiliation{Department of Aeronautics, Imperial College London, South Kensington, London SW7 2AZ.}

\date{\today}

\begin{abstract}
A three-layer asymptotic structure for turbulent pipe flow is proposed, revealing in terms of intermediate variables, the existence of a Reynolds-number invariant logarithmic region. It provides a theoretical foundation for addressing important questions in the scaling of the streamwise mean velocity and variance. The key insight emerging from the analysis is that the scale separation between two adjacent layers is proportional to $\sqrt{Re_{\tau}}$, rather than $Re_{\tau}$. This suggests that, in order to realise Reynolds-number asymptotic invariance,  much higher Reynolds numbers may be necessary to achieve sufficient scale separation.  The formulation provides a theoretical basis for explaining the presence of a power law for the mean velocity in pipe flow at low Reynolds numbers and the co-existence of power and log laws at higher Reynolds numbers. Furthermore, the Townsend-Perry `constant' for the variance is shown to exhibit a systematic Reynolds-number dependence.
\end{abstract}


\maketitle
The drag coefficient of a turbulent boundary layer decreases indefinitely with increasing Reynolds number because the small-scale motion near the surface is always directly affected by viscosity. Reynolds number similarity is therefore an essential tool in the scaling and modelling of near-wall flow.  One of the cornerstones in the theory of turbulent wall flows is the logarithmic (``log'') variation of the mean velocity (Eq. 1) in the inertial sublayer:
\begin{equation} \label{eq1}
U^+ = \frac{1}{\kappa} \mathrm{ln} (y^+) +A,
\end{equation}
where $U^+ = U/u_{\tau}$ and $y^+ = yu_{\tau}/\nu$; $U$ is the mean streamwise velocity, $y$ is the wall-normal distance, $\nu$ is the kinematic viscosity,  $u_{\tau} = \sqrt{\tau_w / \rho}$ is the friction velocity, $\tau_w$ is the wall shear stress and $\rho$ is the density. $\kappa$ in Eq. (1) is the well-known von K\'arm\'an constant.

Another celebrated result in wall turbulence is Townsend's ``attached-eddy'' hypothesis \cite{Townsend76}, which predicts a logarithmic profile for the streamwise (and spanwise) velocity variance in the inertial sublayer. For pipe flows, the log law for streamwise variance takes the form:
\begin{equation} \label{eq2}
\frac {\overline{u^2}} {u_{\tau}^2} = B_1 - A_1 \textrm{ln} \bigg( \frac{y}{R} \bigg),
\end{equation} 
where $u$ is fluctuating streamwise velocity, $R$ is the pipe radius and the overbar indicates time averaging. $A_1$ and $B_1$ are proportionality constants, and $A_1$ is called the Townsend-Perry constant \cite{Marusic_etal2013}. Perry and Chong \cite{Perry_Chong82} showed that the log law for the mean velocity (Eq. 1) and that for the streamwise variance (Eq. 2) can be derived as dual conditions using the attached-eddy formulation.

There remain some central, yet open, questions regarding the Reynolds-number invariance and universality of the von K\'arm\'an constant \cite{Karman30, Millikan38, Townsend76, Marusic_etal2010} that have received much attention, especially for pipe flow \cite{Wosnik_etal2000, Nagib_Chauhan2008, Bailey_etal2014}. By comparison, the Reynolds-number dependence of the Townsend-Perry constant has received rather less attention \cite{Perry_etal86, Hultmark2012, Hultmark_etal2013}. 

There have also been alternative formulations for the mean velocity, e.g. the power-law variation proposed by Barenblatt \cite{Barenblatt93} for pipe and channel flows. Zagarola and Smits \cite{Zagarola_Smits98} used a general matching principle involving different velocity scales for the inner and outer layers, arguing that as long as the ratio of the velocity scales is a function of Reynolds number, the mean velocity is expected to follow a power law. Princeton superpipe measurements show that, at very high Reynolds numbers, $Re_{\tau} = R u_{\tau} / \nu = O(10^5)$, a power law is present in the lower part of the overlap region followed by the log law further away from the wall; see also \cite{McKeon_etal2004, Hultmark_etal2012}.

In this Letter, we propose a theoretical framework, in the context of the turbulent pipe flow, for addressing some of the outstanding issues outlined above. We seek Reynolds-number scaling of the mean velocity and variance in the intermediate region of the pipe flow using the length scale, $y_m^+ \propto \sqrt{Re_{\tau}}$ and the velocity scale ($u_m$) equal to the rms velocity at $y = y_m$. We propose the existence of a distinct intermediate layer (with scales $y_m$ and $u_m$), in addition to the classical inner and outer layers, implying a three-layer asymptotic structure for pipe flow. It should be noted that scaling with $\sqrt{Re_{\tau}}$ is that of the ``meso-layer'', a term that has been used in the literature with different connotations -- either to indicate the location of the peak in the Reynolds shear stress \cite{Sreenivasan_Sahay97} or to provide an offset for the log-law origin in the inertial sublayer \cite{Wosnik_etal2000} or to indicate a region where the turbulent inertia, pressure gradient and viscous forces are in balance \cite{Wei_etal2005}. Here therefore, we use the term ``intermediate layer'', defined as a layer of finite thickness centered on $y/y_m = 1$ with scales $(y_m, u_m)$. This definition is closer in spirit to the intermediate layer proposed by Afzal \cite{Afzal82}. 


The present analysis is based on the NSTAP data measured in the Princeton Superpipe \cite{Hultmark_etal2012}.  Fig. 1 shows scaling of the streamwise variance with length scale, $y_m^+ = 3.5 \sqrt{Re_{\tau}}$ and velocity scale, $u_m = \sqrt{\overline{u^2}} (y=y_m)$: there is an excellent collapse of the profiles in the region around $y / y_m = 1$ for two decades in $Re_\tau$, $1,985 \le Re_\tau \le 98,190$; this is the motivation for using $u_m$ as the intermediate velocity scale.  The choice of constant used in the definition of $y_m^+$ is guided by the coefficients for $\sqrt{Re_{\tau}}$ used in previous definitions of the meso-layer location, e.g. $2\sqrt{Re_{\tau}}$~\cite{Sreenivasan_Sahay97}, or in determining the lower bound of the inertial sublayer, $3\sqrt{Re_{\tau}}$~\cite{Marusic_etal2013}. Here, a slightly higher value of 3.5 is chosen to provide a better $Re_{\tau}$ scaling of the variance profiles for the pipe as well as boundary layer data (not shown). Note that the qualitative (and, to certain extent, quantitative) nature of the results is unaffected by the precise choice of this constant. 
%
 \begin{figure}\label{fig1}
 \includegraphics[width = \columnwidth]{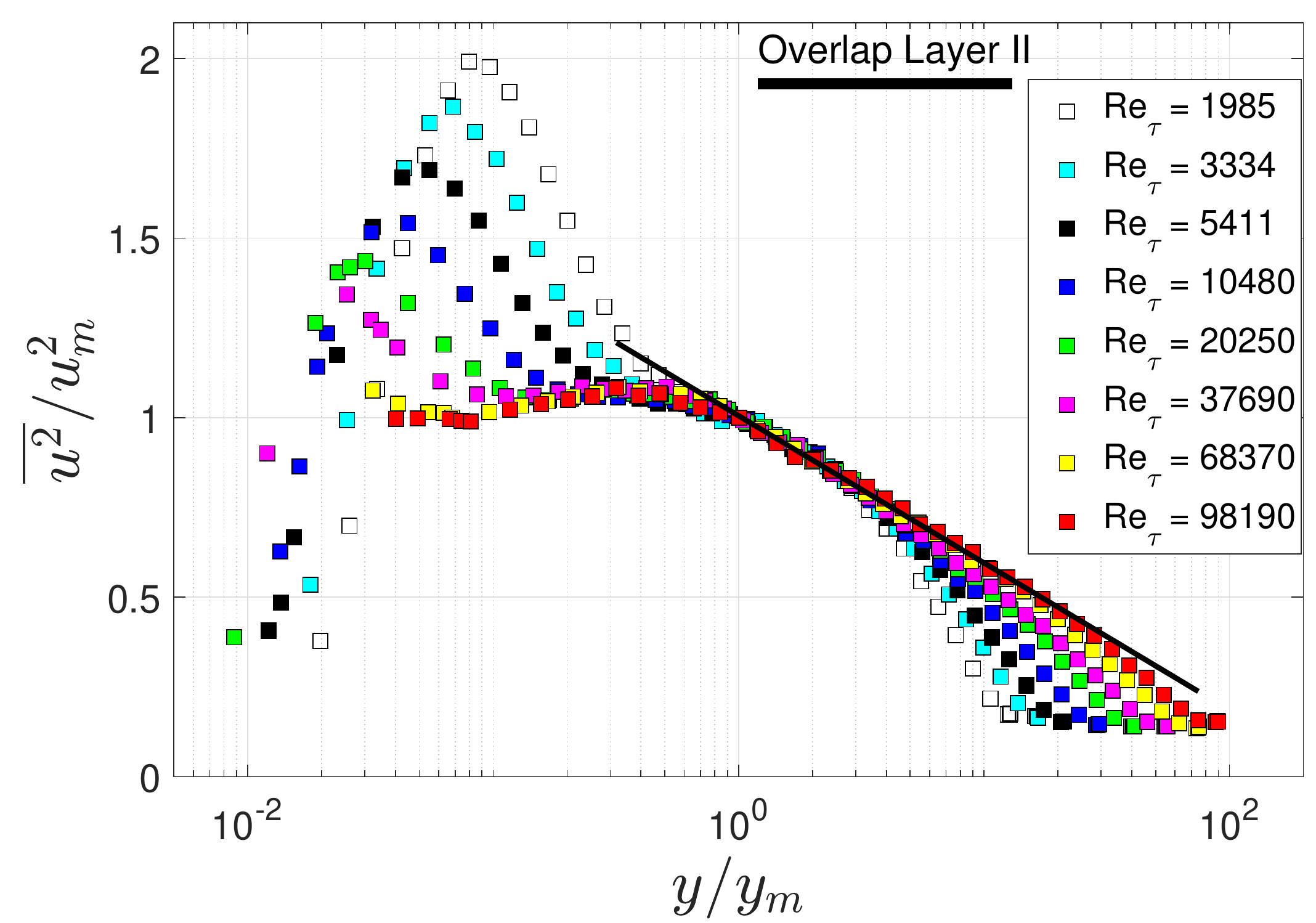}
 \caption{Streamwise variance profiles in a smooth pipe scaled on the intermediate variables $y_m$ and $u_m$; data from Hultmark et al. \cite{Hultmark_etal2012}. The solid line is the log-law fit.}
 \end{figure}

Taking $U_m$ as the mean velocity at $y = y_m$, Fig. 2 shows, in defect form, the corresponding mean velocity profiles scaled on $u_m$.  Excellent scaling is also apparent around $y/y_m = 1$. These scalings suggest the existence of a distinct, asymptotic  intermediate layer lying between the classical inner and outer layers.  This implies that there exists \textit{two} asymptotic overlap regions: one between the inner and intermediate layers (``Overlap Layer I'') and the other between the intermediate and outer layers (``Overlap Layer II'') \cite{Afzal82}.  We choose the velocity scales in the inner and outer layers as $u_i$ and $u_o$ respectively, which are expected to be different from $u_m$. This is in contrast to the earlier formulations \cite{Afzal82,Sreenivasan_Sahay97,Klewicki2013}, which used the same velocity scale, $u_{\tau}$, for all the layers considered. Note also that the five-layer description proposed by Vallikivi et al. \cite{Vallikivi_VGS2015} (based on the spectral characteristics of the streamwise velocity) is different in spirit to the present formulation which, including the two overlap layers, also proposes a total of \textit{five} layers.
 \begin{figure}\label{fig2}
 \includegraphics[width = \columnwidth]{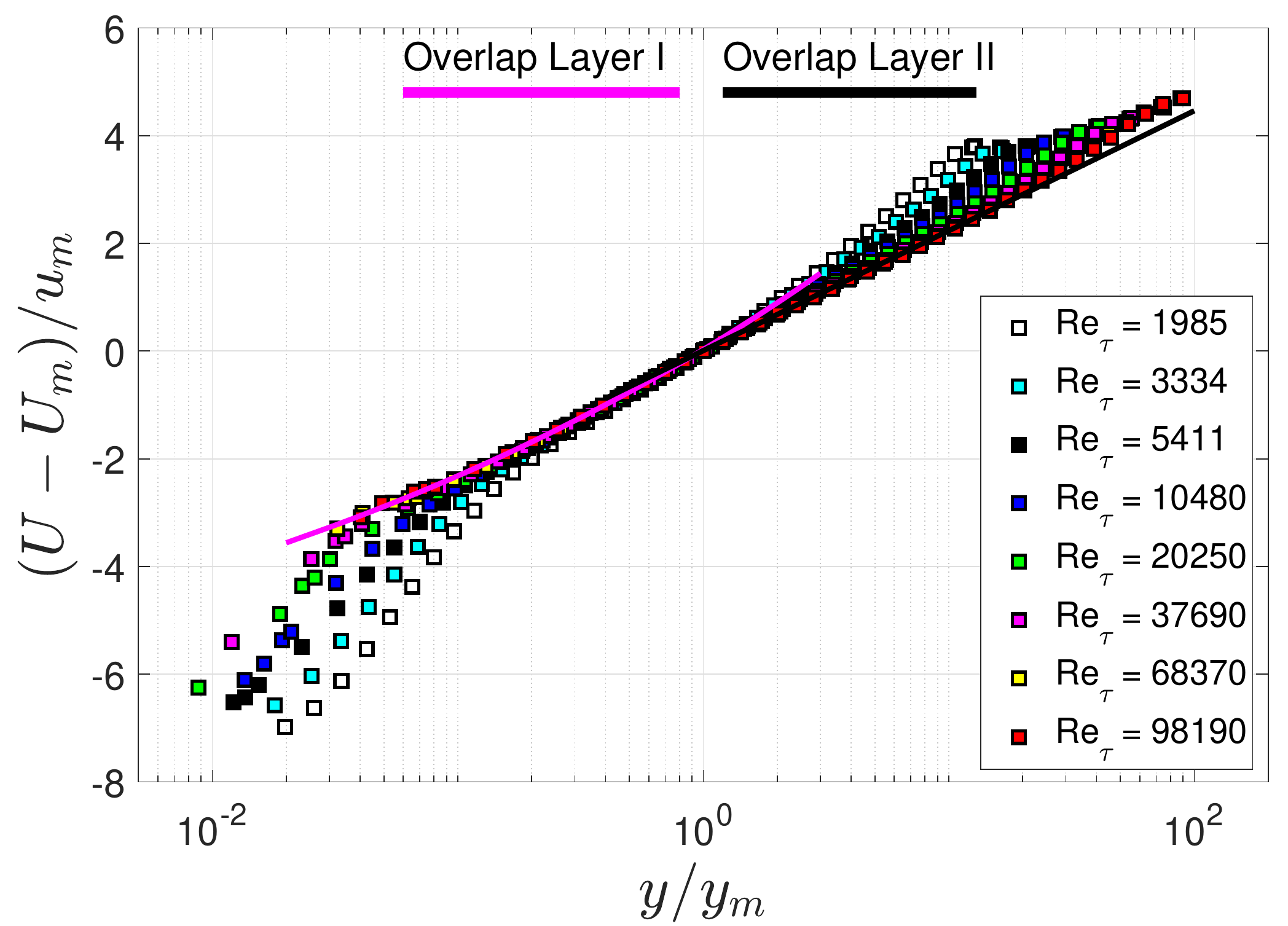}
 \caption{Streamwise mean velocity defect scaled on the intermediate variables. The magenta line indicates a power-law fit and the black line indicates the log-law fit.}
 \end{figure}

For the Overlap Layer I, the inner and intermediate scaling laws are written as,
\begin{equation} 	\label{eq3}
	U^+ = f(y^+), \quad \frac {U-U_m} {u_m} = g(\zeta),
\end{equation}
where $\zeta = y/y_m$. Asymptotic matching of the velocity gradients for the inner and intermediate layers gives:
\begin{equation}	\label{eq4}
	 y^+ f^{\prime} (y^+) = \Lambda_{I} \zeta g^{\prime}(\zeta),
\end{equation}
where $\Lambda_{I} = u_m / u_{\tau}$ and $(^\prime)$ indicates derivatives with respect to the corresponding independent variables. When $\Lambda_I$ is Reynolds-number dependent, Eq. (4) does not imply log-law scaling. Then, Reynolds-number similarity can be achieved by simultaneously matching both velocity and velocity gradient in the overlap region \cite{Zagarola_Smits98}. This results in a power law for the mean velocity: 
\begin{eqnarray} \label{eq5}
  f(y^+) &=& U^+ = C (y^+) ^ \gamma, \\ \nonumber
	\frac{U_m}{u_m} + g(\zeta) &=& \frac{U}{u_m} = C_m \bigg(\frac{y}{y_m} \bigg) ^ \gamma,
\end{eqnarray}
where $\gamma$, $C$ and $C_m$ are constants. Alternatively, when $\Lambda_I$ = constant, a log law is obtained in the overlap region. 

Fig. 3(a) shows the variation of $\Lambda_I$ with Reynolds number, where it continues to increase even at the highest Reynolds number. Therefore, the Overlap Layer I is governed by a power law up to $Re_\tau \approx 10^5$.  Fitting a power-law curve to the $U/u_m$ data for $0.06 \leq y/y_m \leq 0.8$, $Re_{\tau} = 98,190$ gives the power-law constants  as $\gamma = 0.14$ and $C_m = 8.51$. Note that $\gamma$ is independent of $Re_{\tau}$, whereas $C_m$ shows a weak $Re_{\tau}$-dependence in a way that is consistent with $(U - U_m) / u_m$ being Reynolds-number independent. Using these parameters, the variation of $(U - U_m) / u_m$ is plotted in Fig. 2 as a magenta line which fits the data quite well in Overlap Layer I. To determine $C$, we separately fit a power law to the inner-scaled data (not shown here) for $Re_{\tau} = 98,190$ in the corresponding range, $65 \leq y^+ \leq 880$ \citep[see][]{Hultmark_etal2012}. This yields the same value of $\gamma=0.14$, with $C = 8.47$. These values are close to $\gamma = 0.142$ and $C = 8.48$ reported by  \cite{McKeon_etal2004}.
 \begin{figure}\label{fig3}
 \includegraphics[width = \columnwidth]{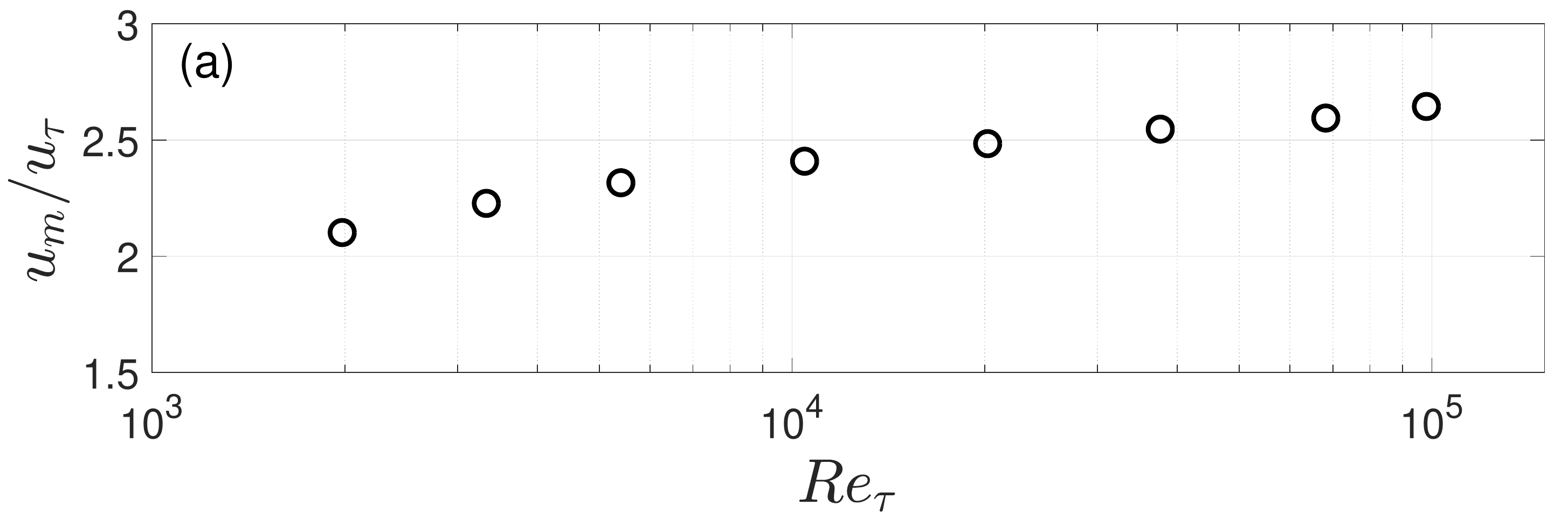}
 \includegraphics[width = \columnwidth]{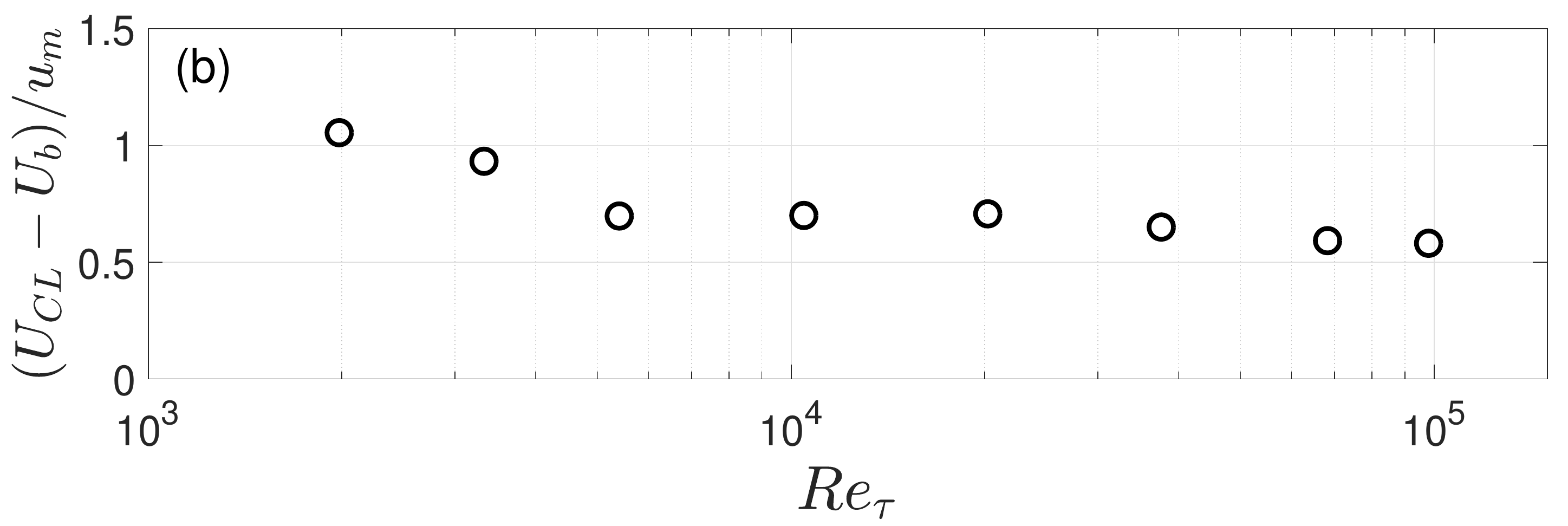}
 \includegraphics[width = \columnwidth]{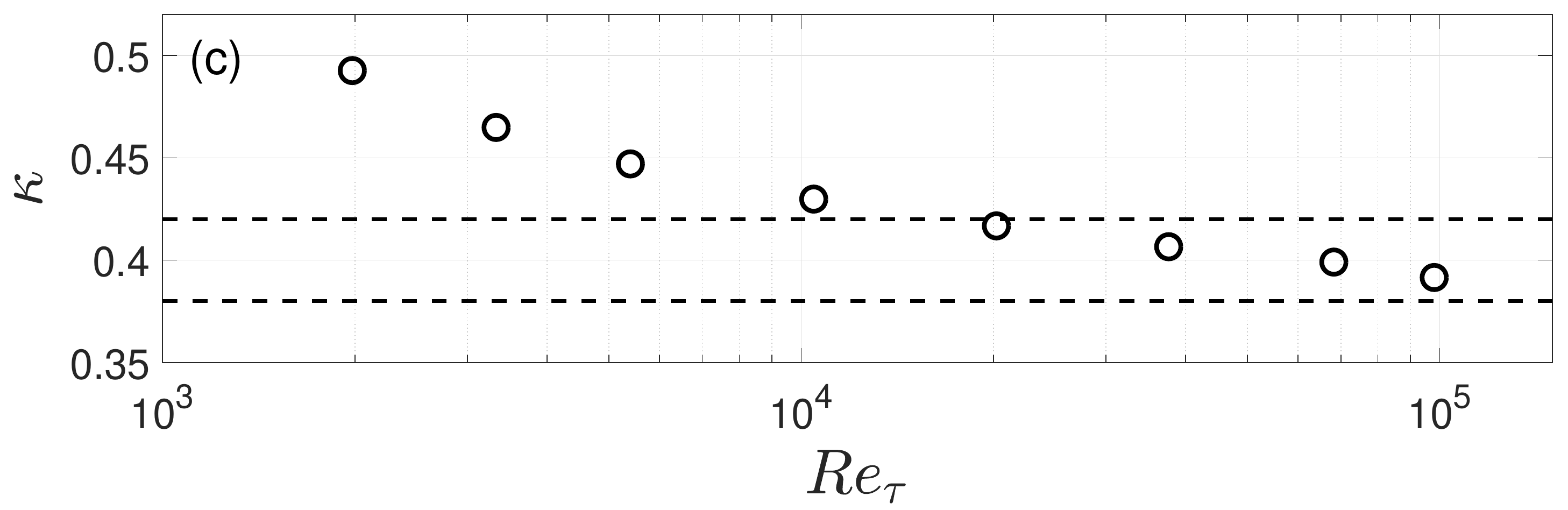}
 \caption{Variation with $Re_{\tau}$ of, (a): $\Lambda_I = u_m / u_{\tau}$, (b): $\Lambda_{II} = (U_{CL} - U_b) / u_m $, (c): $\kappa$ obtained from Eq. (9).}
 \end{figure}

Overlap Layer II is bounded by the intermediate layer and the outer layer. The scaling law for the mean velocity in the outer layer can be written as
\begin{equation} \label{eq6}
	\frac {U_{CL}-U} {u_o} = h\bigg(\frac{y}{R}\bigg),
\end{equation}
where $U_{CL}$ is the pipe centre-line velocity. It can be seen that the character of the mean velocity in Overlap Layer II is determined by the velocity-scale ratio, $\Lambda_{II} = u_o / u_m$ (Eq. 3). Note that although Eq. (6) is written in a Reynolds-number-invariant form, the appropriate velocity scale, $u_o$, that would result in Reynolds-number similarity in the outer region is still unknown \cite{Morrison_etal2004}.

The two alternatives for $u_o$ that have been used so far are $u_{\tau}$ and $U_{CL}-U_b$ \cite{Zagarola_Smits98}, where $U_b$ is the bulk velocity. Choosing $u_o = u_{\tau}$,  $\Lambda_{II} = 1 / \Lambda_{I}$ , implying (Fig. 3a) that Overlap Layer II is also governed by a power law for the entire $Re_{\tau}$ range. This would be a surprising result as there has been overwhelming support in favour of the log law. If, on the other hand, we choose $u_o = (U_{CL} - U_b)$ we get $\Lambda_{II} = (U_{CL} - U_b) / u_m$ -- see Fig. 3(b), which shows that $\Lambda_{II}$ is a strong function of $Re_{\tau}$ for $Re_{\tau} \lesssim 10^4$. Assuming that $\Lambda_{II}$ is approximately constant for $Re_{\tau} > 10^4$,  the log law is recovered for the mean velocity in Overlap Layer II, which, with intermediate variables, can be written as:
\begin{equation} \label{eq7} 
\frac{U - U_m}{u_m} = \frac{1}{\kappa_m} \textrm{ln} \bigg(\frac{y}{y_m}\bigg) + A_m.
\end{equation}

To obtain $\kappa_m$ and $A_m$, we fit a least-square straight line through the mean velocity data (black solid line in Fig. 2) for the two highest Reynolds numbers, $Re_{\tau} = 68,370$ and $Re_{\tau} = 98,190$, and for $1.2 \leq (y/y_m) \leq 13$, equivalent to $4.2\sqrt{Re_{\tau}} \leq y^+ \leq 0.145 Re_{\tau}$ for $Re_{\tau} = 98,190$, which is broadly consistent with the range used in \cite{Marusic_etal2013}. The fit gives the following values for the constants:
\begin{equation} \label{eq8} 
\kappa_m = 1.034 \quad \quad A_m = 0.0084.
\end{equation}
Note that, provided the Reynolds-number similarity in the intermediate layer is ensured, the value of $\kappa_m$ is independent of the choice of the coefficient in the definition of $y_m$.  The value of $A_m$, however, depends on this choice (Eq. 7). The classical log-law constants can be expressed in terms of $\kappa_m$ and $A_m$ (Eqs. 1,7) as:
\begin{eqnarray} \label{eq9}
\kappa &=& \frac{\kappa_m} {(u_m / u_{\tau})}, \\ \nonumber
A &=& \frac{u_m}{u_{\tau}} \bigg\{\bigg[ \frac{U_m}{u_m} + A_m \bigg] - \frac{1}{\kappa_m} \textrm{ln} (y_m^+) \bigg\}.
 \end{eqnarray}
$\kappa$ obtained from Eq. (9) is plotted in Fig. 3(c) and shows a systematic decrease with Reynolds number; see also table 1. For $Re_{\tau} > 10^4$, $\kappa$ falls within the uncertainty band of $0.4 \pm 0.02$ \cite{Bailey_etal2014}, shown as dashed lines in the figure; the trend exhibited by $\kappa$ within the band can be traced back to the weak variation in $\Lambda_{II}$ for $Re_{\tau} > 10^4$ (Fig. 3b). For $Re_{\tau} < 10^4$, the values of $\kappa$ (and $A$; table 1) are much higher than those which could be reasonably associated with the log law. This suggests that the mean velocity profile in Overlap Layer II should really follow a power law for these lower Reynolds numbers, as also implied by the strong Reynolds-number dependence of $\Lambda_{II}$ for $Re_{\tau} < 10^4$ (Fig. 3b). Moreover, at these Reynolds numbers the two overlap layers may not be entirely distinct and therefore the two power laws may appear indistinguishable (Fig. 2). These results are consistent with the observations in \cite{Zagarola_Smits98} and \cite{McKeon_etal2004}.

The presence of a power law in Overlap Layer I and of the log law in Overlap Layer II, for $Re_{\tau} > 10^4$, supports the observation by Zagarola and Smits \cite{Zagarola_Smits98} \citep[see also][]{Hultmark_etal2012} that, at high $Re_{\tau}$, the mean velocity initially follows a power-law profile followed by the log-law variation. While in these analyses the power and log laws share the same overlap region, in the present three-layer formulation they occupy two different overlap regions. This provides an explanation for the co-existence of the power and log-law profiles in the pipe flow at a given (and sufficiently large) Reynolds number. Furthermore, since the length scale for the intermediate layer is $\propto \sqrt{Re_{\tau}}$, the lower limit for the log law for the mean velocity should be Reynolds-number dependent rather than constant in wall variables. \citep[See][]{Marusic_etal2013,Bailey_etal2014}.

\vspace{-0.4cm}
\begin{table}[h]   
  \centering
	\label{tab1}
	 \caption{Variation of the log-law constants for the mean velocity ($\kappa$ and $A$) and variance ($A_1$ and $B_1$) with $Re_{\tau}$. These are obtained from Eqs. (9) and (11) using log fits (Eqs. 7 and 10) in the range $1.2 \leq y/y_m \leq 13$.}
  \begin{tabular}{ccccc}
				$Re_{\tau}$ & \quad $\kappa$ & \quad $A$ &	\quad	$A_1$	& \quad   $B_1$ \\
				\hline
				 1,985  & \quad 0.492   & \quad 7.04 &  \quad 0.785 & \quad 2.435 \\
				 3,334  & \quad 0.465  & \quad 6.554  &  \quad 0.882 & \quad 2.506 \\
				 5,411  & \quad 0.447  & \quad 5.99  &  \quad 0.953 & \quad 2.479 \\
				10,480  & \quad 0.43   & \quad 5.571 &  \quad 1.031 & \quad 2.341 \\
				20,250  & \quad 0.417  & \quad 5.171 &  \quad 1.097 & \quad 2.129 \\
				37,690  & \quad 0.406  & \quad 4.919 &  \quad 1.153 & \quad 1.879 \\
				68,370  & \quad 0.399  & \quad 4.747 &  \quad 1.197 & \quad 1.594 \\
				98,190  & \quad 0.391  & \quad 4.545  &  \quad 1.243 & \quad 1.431 \\				         				
  \end{tabular}
\end{table}

In Fig. 1, the Reynolds-number similarity of the streamwise variance for $y \approx y_m$ implies that, for $y > y_m$ there should exist a Reynolds-number-invariant log law scaled on the intermediate variables, and given as:
\begin{equation} \label{eq10} 
\frac {\overline{u^2}} {u_{m}^2} = B^m_1 - A^m_1 \textrm{ln} \bigg( \frac{y}{y_m} \bigg).
\end{equation} 
To determine $A^m_1$ and $B^m_1$, we fit a least-square straight line through the points in Fig. 1 (shown as a solid line) in the region $1.2 \leq y/y_m \leq 13$ for the two highest Reynolds numbers; this range is the same as that chosen for fitting a log law for the mean velocity in Overlap Layer II (Fig. 2). (The behaviour of $\overline{u^2}$ in Ovelap Layer I is beyond the scope of the present work.) This gives $A^m_1 = 0.178$ and $B^m_1 = 1.005$. We do not attempt to estimate the uncertainty bounds for $A^m_1$ and $B^m_1$ (and also for $\kappa_m$ and $A_m$; Eq. 8) here, as their precise numerical values are not relevant for the key conclusions of the paper. The classical constants, $A_1$ and $B_1$ (Eq. 2), can be readily expressed in terms of $A^m_1$ and $B^m_1$ as
\begin{eqnarray} \label{eq11}
 A_1  &=& A_1^m \bigg( \frac {u_m^2} {u_{\tau}^2} \bigg), \\ \nonumber
 B_1  &=& \bigg[A_1^m \textrm{ln} \bigg(\frac {y_m} {R} \bigg) + B_1^m \bigg] \bigg( \frac {u_m^2} {u_{\tau}^2} \bigg).
\end{eqnarray}

 \begin{figure}[t!]\label{fig4}
 \includegraphics[width = \columnwidth]{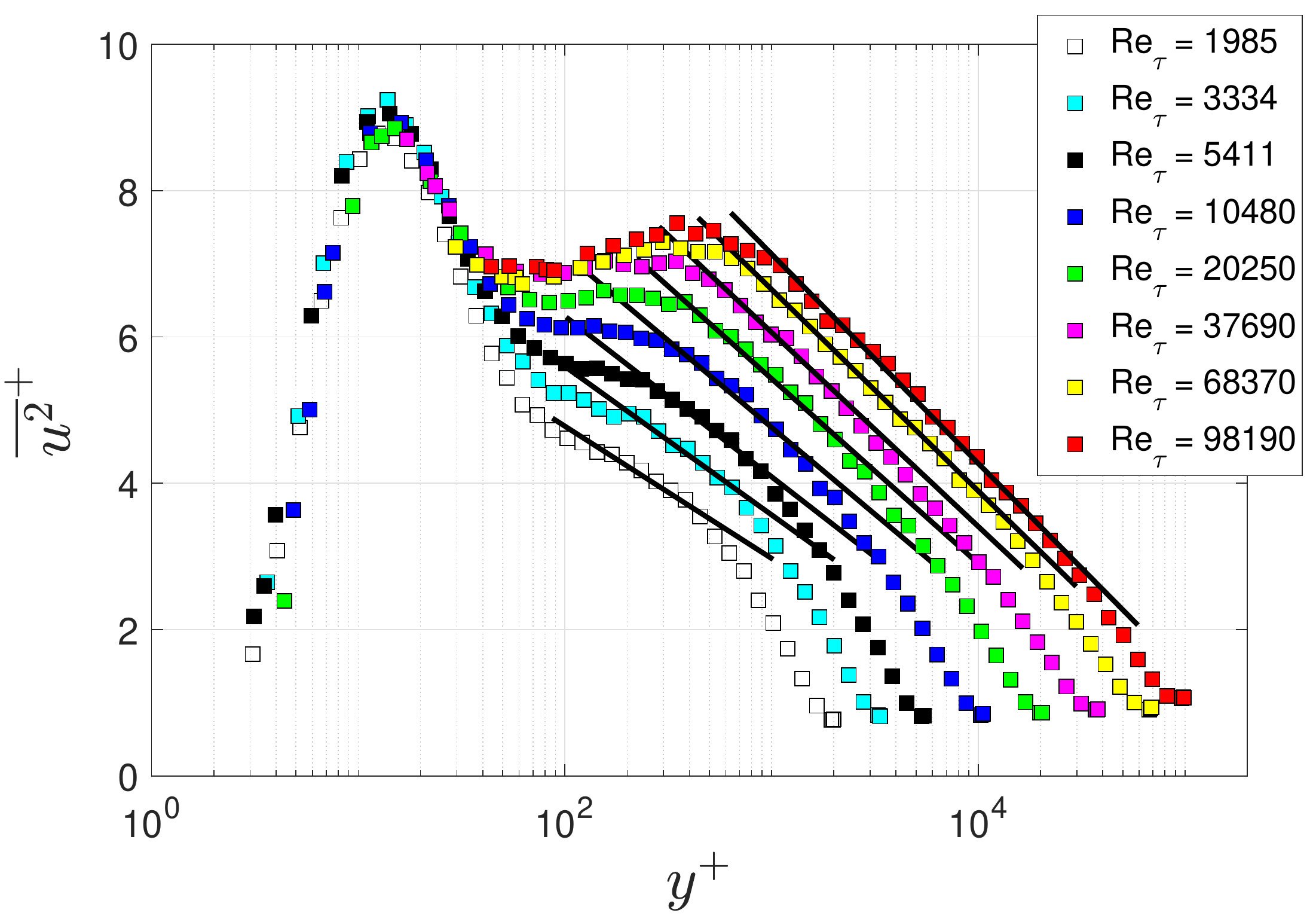}
 \caption{Streamwise variance profiles for the pipe; solid lines are the classical log-law fits using $A_1$ and $B_1$ from table 1.}
 \end{figure}
 
$A_1$ and $B_1$ calculated from Eq. (11) (with $A^m_1 = 0.178$ and $B^m_1 = 1.005$) are included in table 1; a clear trend in $A_1$ and $B_1$ with respect to $Re_{\tau}$ is evident. Fig. 4 shows the log-law fits to the variance, in wall variables, obtained by using $A_1$ and $B_1$ from table 1. As can be seen, the log fits inferred from Eq. (11) show a good match with the measured profiles in the intermediate region, over the entire $Re_{\tau}$ range. This leads us to conclude that the Townsend-Perry `constant', $A_1$, actually shows a systematic dependence on $Re_{\tau}$ even for $Re_{\tau} > 2 \times 10^4$. This is due to the fact that $A^m_1$ and $B^m_1$ are Reynolds-number invariant and that $u_m/u_{\tau}$ (Fig.  3a) and $y_m/R \; (=3.5/\sqrt{Re_{\tau}})$ show a continuous dependence on $Re_{\tau}$. Furthermore, the values of $A_1$ in table 1 are entirely consistent, at corresponding Reynolds numbers, with those in Perry et al. \cite{Perry_etal86} ($A_1 = 0.9$ for $Re_{\tau} \leq 3,900$) and Hultmark et al. \cite{Hultmark_etal2012} ($A_1 = 1.25$ for $Re_{\tau} = 98,190$; Marusic et al. \cite{Marusic_etal2013} reported $A_1 = 1.23\pm0.05$ for the same $Re_{\tau}$), and provide an explanation for the observation. 

%
We are grateful to Professor Lex Smits for use of the NSTAP data.  We acknowledge financial support from EPSRC under Grant No. EP/I037938/1.

\bibliographystyle{apsrev4-1} 
\bibliography{References_PRL_scaling}

\end{document}